\documentclass[pra,twocolumn,showpacs,floatfix]{revtex4}

\usepackage{amsmath}
\usepackage{times}
\usepackage{txfonts}
\usepackage{graphicx}
\usepackage{color}
\usepackage{float}
\usepackage[colorlinks,linkcolor=blue,citecolor=blue,urlcolor=blue]{hyperref}
\graphicspath{Figures/}  

\usepackage{vector}  
\hypersetup{urlcolor=blue, colorlinks=true}  

\begin{document}

\title{Stationary and traveling solitons via local dissipations \\
in Bose-Einstein condensates in ring optical lattices} 

\author{Russell \surname{Campbell}}
\author{Gian-Luca \surname{Oppo}} 
\affiliation{Institute of Complex Systems, SUPA and Department of Physics, 
University of Strathclyde, 107 Rottenrow, Glasgow G4 0NG, Scotland, UK.}
\email{russell.campbell@strath.ac.uk}  

\begin{abstract}
A model of a Bose-Einstein condensate in a ring optical lattice with atomic dissipations applied at a stationary or at a moving location on the ring is presented.  The localized dissipation is shown to generate and stabilize both stationary and traveling lattice solitons.  Among many localized solutions, we have generated spatially stationary quasiperiodic lattice soltions and a family of traveling lattice solitons with two intensity peaks per potential well with no counterpart in the discrete case.  Collisions between traveling and stationary lattice solitons as well as between two traveling lattice solitons display a critical dependence from the lattice depth. Stable counterpropagating solitons in ring lattices can find applications in gyroscope interferometers with ultra-cold gases.
\end{abstract}


\maketitle

\section{Introduction}

Bose-Einstein condensates (BEC) trapped in an optical lattice have attracted a major scientific interest and can provide an interesting analogue to solid-state systems \cite{morsch06,bloch08}.  An advantage here is that there is almost complete control of the parameters that regulate the lattice.  This has led to studies of solid-state phenomena such as quantum phase transitions \cite{greiner02}, transport \cite{chiofalo01}, Anderson localization \cite{roati08} and macroscopic Zeno effect \cite{zezyulin12} . In the superfluid phase of the BEC, a lot of attention has been devoted to discrete breathers in the discrete nonlinear Schr\"odinger equation (DNLSE) \cite{franzosi11} and to lattice solitons in the Gross-Pitaevskii equation  (GPE) \cite{strecker02}.  
The optical lattice allows solitons and discrete breathers to exist with repulsive BEC where they have been observed experimentally \cite{eiermann04}.  
Methods for the generation of discrete breathers include the evolution from Gaussian wavepackets \cite{trombettoni01,hennig16} and the relaxation from random phase states via localized losses \cite{livi06}.  Stabilization of discrete breathers in the DNLSE via localized losses can be acheived by either the progressive lowering of the fluctuating background \cite{livi06,franzosi11} or by producing sudden atomic avalanches \cite{ng09}.  Moving discrete breathers have also been obtained with these techniques in accurate numerical simulations. An interesting application of moving breathers is in atom interferometry \cite{franzosi07}.  Without a lattice, methods of soliton interferometry have been implemented experimentally in \cite{mcdonald14} while techniques for generating counterpropagating solitons by using a splitting potential barrier in a ring trap have been proposed and discussed in \cite{helm12,helm14,helm15}. The aim of our work is to demonstrate that stationary and moving lattice solitons in continuous models of BEC in ring lattices can be generated and stabilized via localized losses. In particular we show that higher order lattice solitons that have no counterpart in the discrete case can be effectively stabilized by these techniques.

\begin{figure}[htb!]
\centering
\includegraphics[angle=0.0,clip,width=0.8\columnwidth]{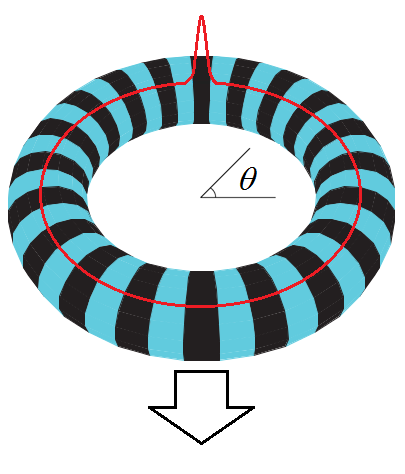}
\caption{\color{black}(Color online) Lattice soliton in an optical lattice ring-trap. 
  \label{fig:ring}}
\end{figure}

We consider a ring trap \cite{heathcote08} with a toroidal optical lattice as realized for example in \cite{amico05,frankearnold07,henderson09,moulder12} (see Fig.~\ref{fig:ring}). Experimentally, a BEC in a ring trap with an azimuthal optical lattice can be achieved by either using counter-propagating laser beams in a circular wave-guide or by illuminating transversally a ring trap with two counter-rotating orbital angular momentum laser beams with optical axis along the centre of the ring trap and perpendicular to the trap.  It is important to outline that the equations used in this paper for the case of a BEC in an optical lattice also describe light traveling through a cylindrical array of optical waveguides. All the results presented here can then be extended to this purely optical case.

Model equations for a BEC in an ring optical lattice are introduced in Section \ref{model}. These are the continuous counterpart of the DNLSE with the addition of localised dissipations. In order to differentiate and compare the solutions of the continuous model of Section \ref{model} with those of the DNLSE, we refer to continuous soliton solutions in the annular periodic potential as lattice instead of discrete solitons. Lattice solitons are also known as 'gap solitons' in the literature. In Section \ref{stationary} we discuss the generation of symmetric and asymmetric lattice solitons via the effect of stationary localized losses and compare them succesfully to those found by other numerical methods in \cite{efremidis03}. Traveling lattice solitons (TLS) in the ring trap are generated and investigated in Section \ref{travelling}. Two kind of TLS are found: with one peak per lattice well and with two peaks per lattice well. It is important to note that the double peak TLS has no counterpart in the discrete NLS. Finally, collisions between traveling and stationary lattice solitons in a ring trap are investigated in Section \ref{collisionTSb}, while collisions between two traveling lattice solitons are studied in Section \ref{collisionTT}.  Possible applications to atom interferometry are discussed in the conclusions.

\section{The model equations}
\label{model}

We consider the Gross-Pitaevskii equation for a one dimensional BEC in an optical lattice given by \cite{efremidis03,louis03}:  
\begin{equation}
  i \hbar \frac{\partial \Psi(x,T )}{\partial T} = \left(-\frac{\hbar}{2m}\frac{\partial^2}{\partial x^2} + E_0 sin^2 \left(\frac{\pi x}{L}\right) 
  + g_{1D} |\Psi|^2 \right) \Psi \, ,
  \label{GPE1D}
\end{equation}
where $\hbar$ is the reduced Planck's constant, $E_0$ is the potential depth (usually measured with respect to the recoil energy), $L=\lambda/2$ is the lattice period, $\lambda$ the laser or spatial wavelength used for the optical lattice and $m$ the atomic mass. The one-dimensional atom-atom interaction parameter is given by $g_{1D}=2\hbar\omega_\perp a_s$, where $\omega_{\perp}$ is the transverse trapping frequency and $a_s$ the scattering length of the BEC.

To describe the BEC trapped in the ring we use Eq. (\ref{GPE1D}) provided with periodic boundary conditions. For convenience, dimensionless variables are used. First we rewrite Eq.~(\ref{GPE1D}) by normalizing $u=\sqrt{L/2N}\Psi$, $t=T/T_0$ and $V_0=E_0/E_r$, where $T_0=mL^2/4\hbar$, $E_r=4\hbar^2/mL^2$ is the recoil energy and $N$ is the number of atoms \cite{efremidis03}. The length scale $x$ is then changed into the ring angle $\theta=2 \pi x/M L$ ranging from 0 to $2\pi$ radians, where $M$ is the number of potential wells in the ring along the azymuthal direction. The resulting equation is: 
\begin{equation}
  i \frac{\partial u(\theta,t )}{\partial t} = \left(-\frac{\pi^2}{2M^2}\frac{\partial^2}{\partial \theta^2} + V_0 sin^2 \left(\frac{M \theta}{2}\right) 
  + \beta |u|^2 - i \rho(\theta,t) \right) u \, .
  \label{GPEnorm}
\end{equation}
The nonlinear parameter $\beta=N \omega_\perp a_s m L/\hbar$ is positive for repulsive condensates and negative for attractive ones. In order to describe localized losses of the atomic population along the ring at certain times $t$, we have added the term $- i \rho(\theta,t) u$ in Eq. (\ref{GPEnorm}). Extremely precise methods for removing atoms in a particular position of a BEC in optical lattices have been implemented with the use of narrow electron beams \cite{gericke08}. The intensity of such electron beams can control the number of atoms that are removed from one or more potential wells of the optical lattice.  In our examples here, localized losses are applied at the furthest point in the ring (i.e. at an angular distance of $\pi$ radians) from the peak of the stationary or moving lattice soliton. For example, with the stationary lattice solitons that are usually generated at $\theta=\pi$, the dissipation is applied at $\theta=0=2\pi$.

Equation \ref{GPEnorm} is normalized so that at $t=0$, before any atoms are lost due to dissipation,
\begin{equation}
 \int u(t=0) \; d\theta = 1 \, .
  \label{GPEint}
\end{equation}

\section{Stationary Localized Dissipations}
\label{stationary}

Stationary and moving breathers can be formed in the DNLS starting from initial Gaussian wavepackets \cite{hennig16,trombettoni01,franzosi11}. For our continuous variable model, we use the general form:

 \begin{equation}
  u(t=0) = \frac{M^2}{\gamma^{1/2}\pi^{9/4}}
  \exp{\left(-\frac{(\theta-\theta_c)^2}{2\gamma^2}\right)} 
  \label{gaussinit}
\end{equation}

with $\theta_c$ being the position of the centre of the wave-packet and $\gamma$ the width. With the nonlinear coefficient fixed at $\beta=1$, the initial width was changed and several localized solutions were found in the case of zero losses (i.e. the conservative case). 

Typically, the Gaussian wavepacket would reshape into a solitonic profile. The atomic mass expelled from the wavepacket, however, forms a noisy backround. The peak fluctuates in height as it keeps interacting with the background. As the width of the initial wave-packet is increased, the background becomes noisier and sometimes smaller amplitude peaks appear close to the main one. The small amplitude peaks, however, do not survive in the long term. When the width of the initial Gaussian condition is too large, no peak is formed and the condensate disperses onto the background. Similarly, if the width is too small (smaller than a single potential well), there is no self-localization either.

\begin{figure}[htb!]
 \includegraphics[angle=0.0,clip,width=0.8\columnwidth]{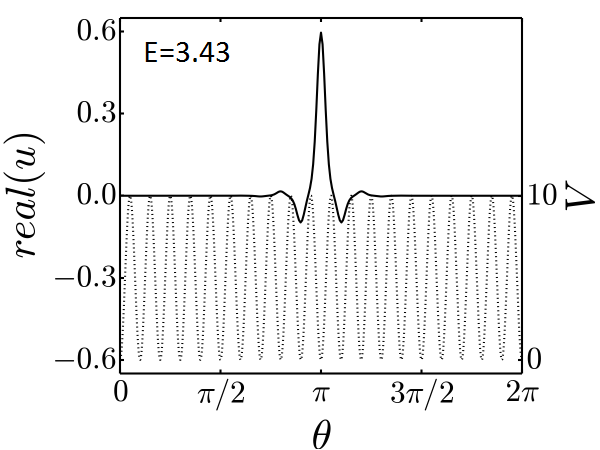}
\caption{Stationary lattice soliton formed from applying dissipation to an initial Gaussian 
wavepacket (\ref{gaussinit}). The shape of the lattice soliton is very similar to those presented in \cite{efremidis03}. The dotted line is the lattice $V$, see the scale on the right, with $V_0=10$.
\label{figuregausscon} }
\end{figure}

When dissipation is applied to the above configuration, we obtain less noisy backgrounds since the mass expelled from the initial wavepacket escapes at the location of the losses. In all the examples in this section, the dissipation acts on around 4 potential wells with the maximum loss of 0.5 at $\theta=0=2\pi$. For a Gaussian of unit width ($\gamma=1$) we routinely recover stable lattice soliton solutions via localized dissipation (see, for example Fig.~\ref{figuregausscon}). These solutions are very close to those shown in \cite{efremidis03} and obtained with very different numerical methods. 

\begin{figure}[htb!]
\includegraphics[angle=0.0,clip,width=\columnwidth]{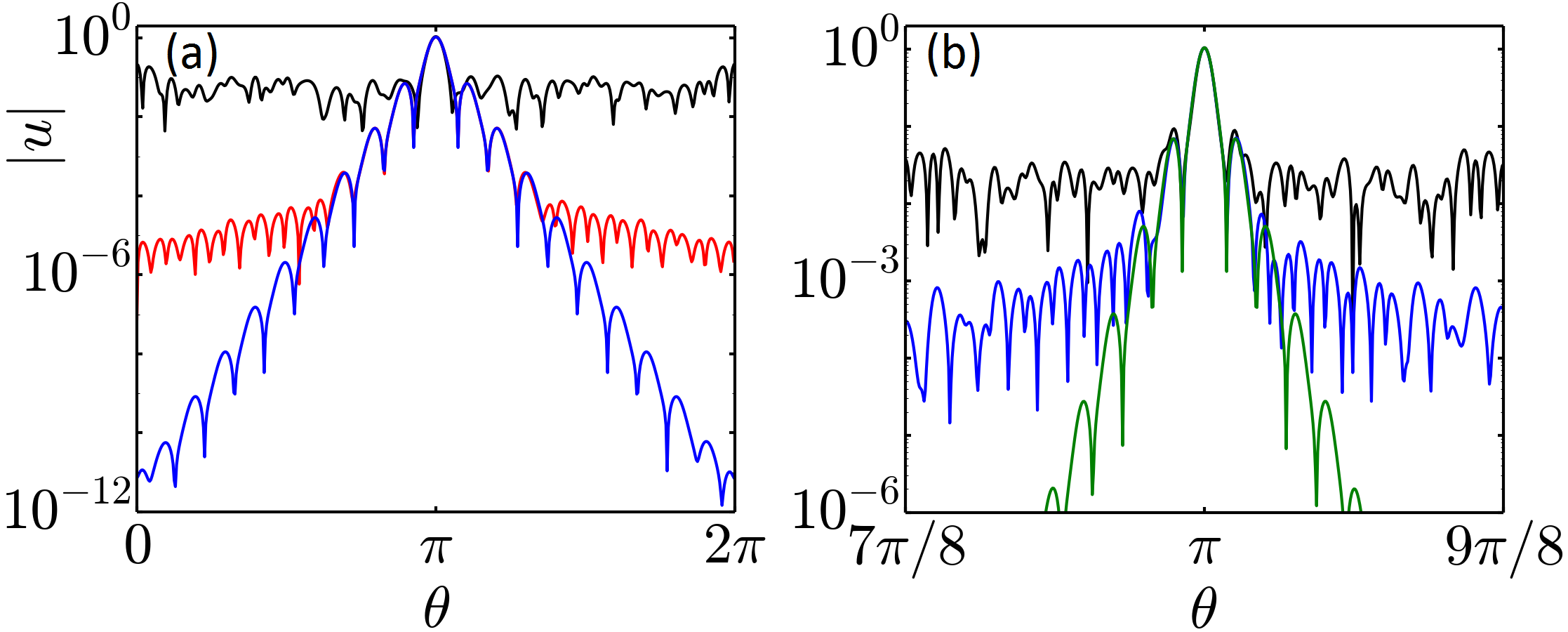}
\caption{\color{black}(Color online) Intensity distribution of stationary lattice solitons obtained from initial Gaussian wavepackets with localized dissipation. The curves correspond to t=0 (black), t=20000 (red), and t=100000 (blue) in a lattice of 20 potential wells (a) and 160 potential wells (b). Time t=0 corresponds to the moment dissipation is turned on. In (b), we also show the intensity distribution after increasing the number of potential wells where the dissipation acts on from $\sim 4$ to $\sim 150$ at $t=100000$ and then running the simulation for another 100000 time units.
\label{figurediss}}
\end{figure}

The effect of dissipation on the soliton and background can be seen clearly in Fig.~\ref{figurediss}, which shows the decay of the noisy backround leading to exponential tails associated with lattice solitons. In the larger lattice, this effect is less obvious, due to the distance from the lattice soliton to the place where the dissipation is applied (Fig.~\ref{figurediss} (b)). Making the dissipation broader so that it acts on most of the potential wells in the lattice (in this case $\sim150$ out of 160) can help to reveal the tails faster (see green line in Fig.~\ref{figurediss} (b)).

  \begin{table}
    {\color{black}
    \caption{Values of parameters used in the simulations \label{parametertable}}
    \begin{tabular}{c c c c c c c c}
      \hline
      \hline
      $\gamma$ & Peak Intensity & Gradient of tails & Frequency &Nature of Solution\\
      \hline
      0.5 & 1.258 & 0.420 & 4.05 & SLS\\
      0.7 & 1.167 & 0.410 & 3.97 & SLS \\
      0.9 & 1.012 & 0.398 & 3.89 & SLS \\
      1.0 & 0.923 & 0.392 & 3.83 & SLS\\
      1.2 & 0.765 & 0.367 & 3.72 & SLS\\
      1.3 & 0.651 & 0.355 & 3.62 & QS\\
      1.6 & 0.539 & 0.332 & 3.55 & SLS \\
      1.8 & 0.377 & 0.289 & 3.46 & QS\\
      2.0 & 0.362 & 0.287 & 3.43 & SLS\\
      \hline
    \end{tabular}}
  \end{table}

We find that the final shape and frequency of the lattice soliton is affected by changing the initial width of the Gaussian $\gamma$: the wider is the Gaussian, the more atoms are lost due to dissipation and the lower is the final peak amplitude of the lattice soliton. The frequency of the oscillations of the real/imaginary parts of the solitons, along with the gradient of the exponential tails of the soliton, is larger if the number of atoms (i.e. the peak amplitude) is larger. This can be seen in Table ~\ref{parametertable} where the peak intensity, gradient and the frequency of the final lattice soliton are displayed versus the Gaussian width. 

\begin{figure}[htb!]
\includegraphics[angle=0.0,clip,width=\columnwidth]{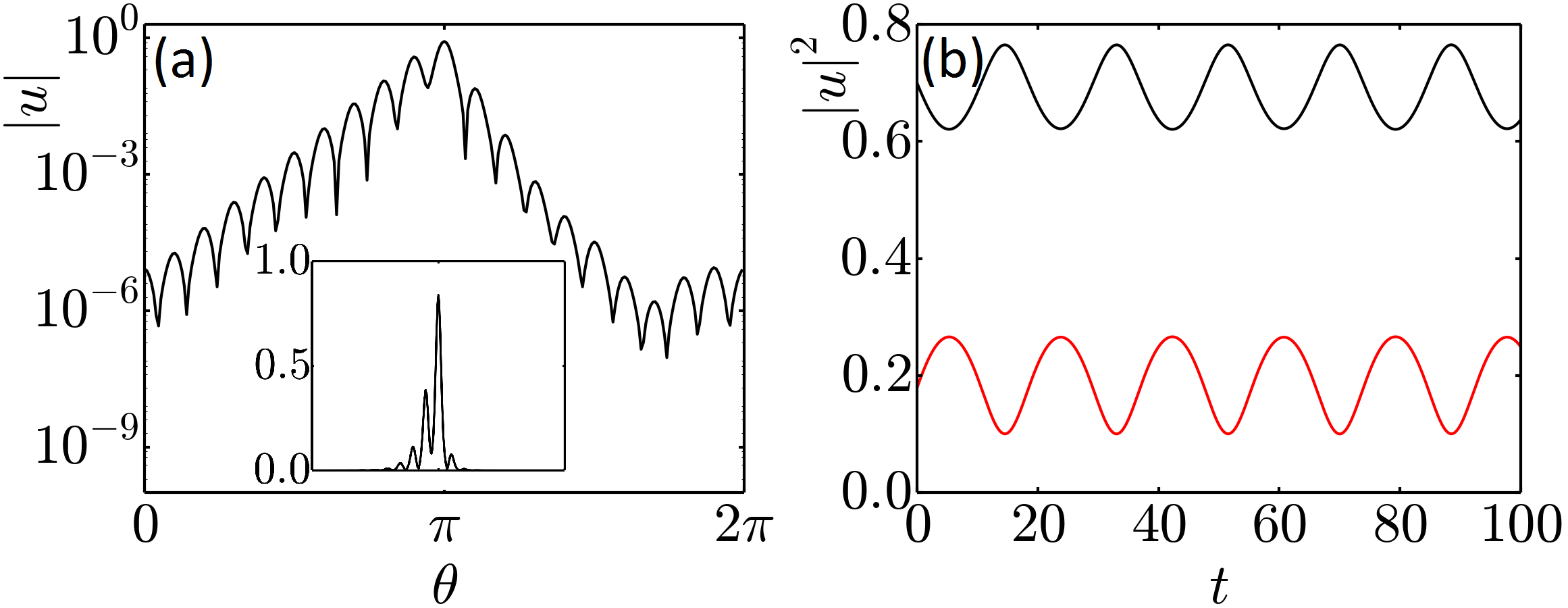}
\caption{\color{black}(Color online) (a) Quasiperiodic solution generated by applying dissipation to an initial Gaussian with width $\gamma=1.3$. (b) Variation in time of intensity of peak of larger amplitiude (black line) and smaller amplitude (red line).
  \label{quasi}}
\end{figure}

Localized dissipations allow one to generate a broad variety of lattice solitons from Eq. (\ref{GPEnorm}). For example for $\gamma=1.3$ and $\gamma=1.8$, the result is that of asymmetric lattice solitons with two high-peaks next to each other (see Fig. \ref{quasi}). The oscillation of this asymmeteric solution is quasiperiodic.  The values of peak intensity, frequency and gradient of the tails of the quasiperiodic solutions (QS) in Table ~\ref{parametertable} are those associated with the highest peak in each case. Note that there are quasiperiodic discrete breather counterparts in the DNLSE (see \cite{johansson97}).

\begin{figure}[htb!]
\centering
\includegraphics[angle=0.0,clip,width=0.80\columnwidth]{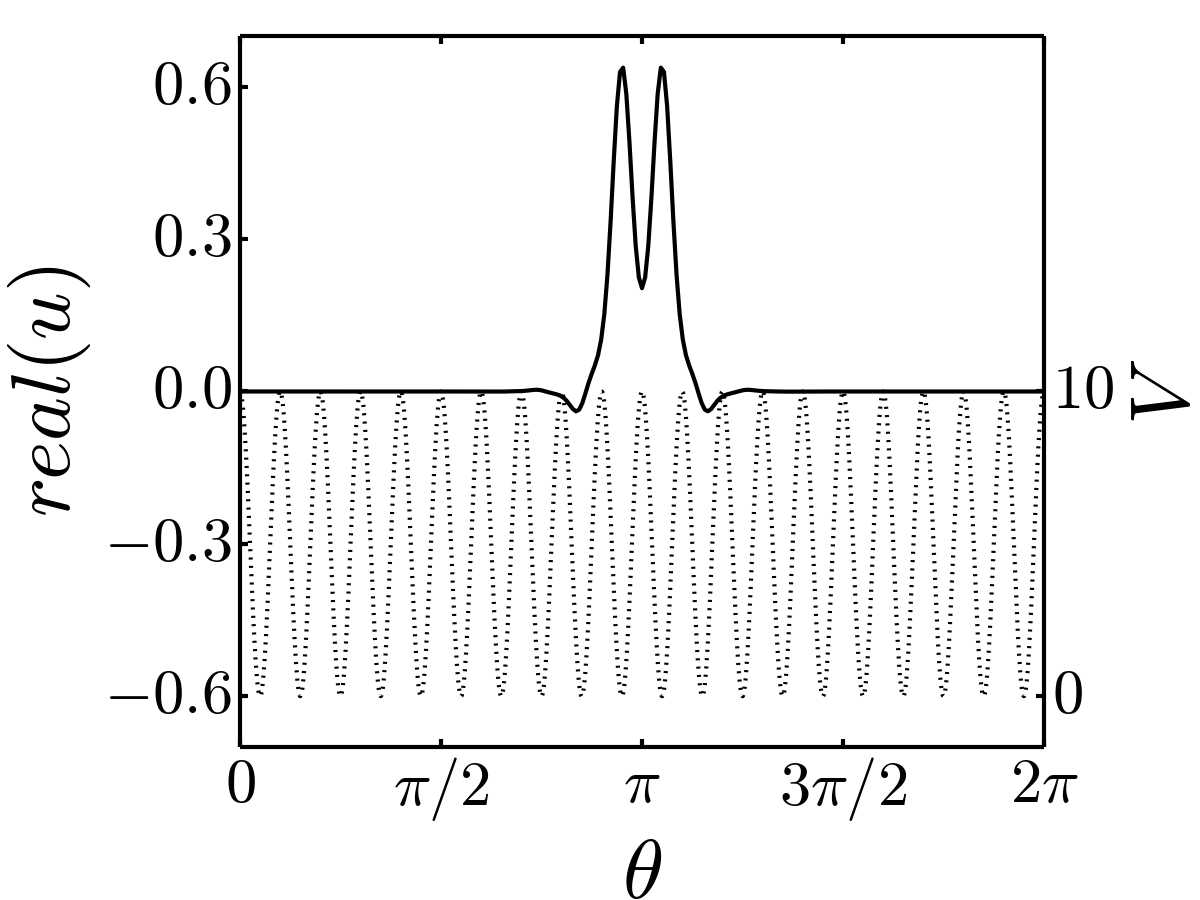}
\caption{\color{black}(Color online) A higher-order stationary soliton solution with two-peaks formed from applying dissipation to an initial Gaussian wavepacket (\ref{gaussinit}) centered between two potential wells.
  \label{twin}}
\end{figure}

Another type of solution, shown in Fig. \ref{twin}, is symmetric with two main peaks (as in \cite{louis03}). The two peaks are in-phase with each other and oscillate at the same frequency, as opposed to the previous quasiperiodic example in Fig. \ref{quasi}. This lattice soliton has been found by using localized dissipations and by shifting the initial wavepacket by $L/2$ (half a potential well). The same effect can be obtained  with a potential of $V=V_0 cos^2 (M \theta/2)$ rather than $V=V_0 sin^2 (M \theta/2)$, so that the initial Gaussian wavepacket is centered between two potential wells. The nonlinearity is set to the value of $\beta=10$, corresponding to a higher number of initial atoms or a larger scattering length. With $\beta=1$, the double peak relaxes to the single peak solution quickly.

\begin{figure}[htb!]
\centering
\includegraphics[angle=0.0,clip,width=0.8\columnwidth]{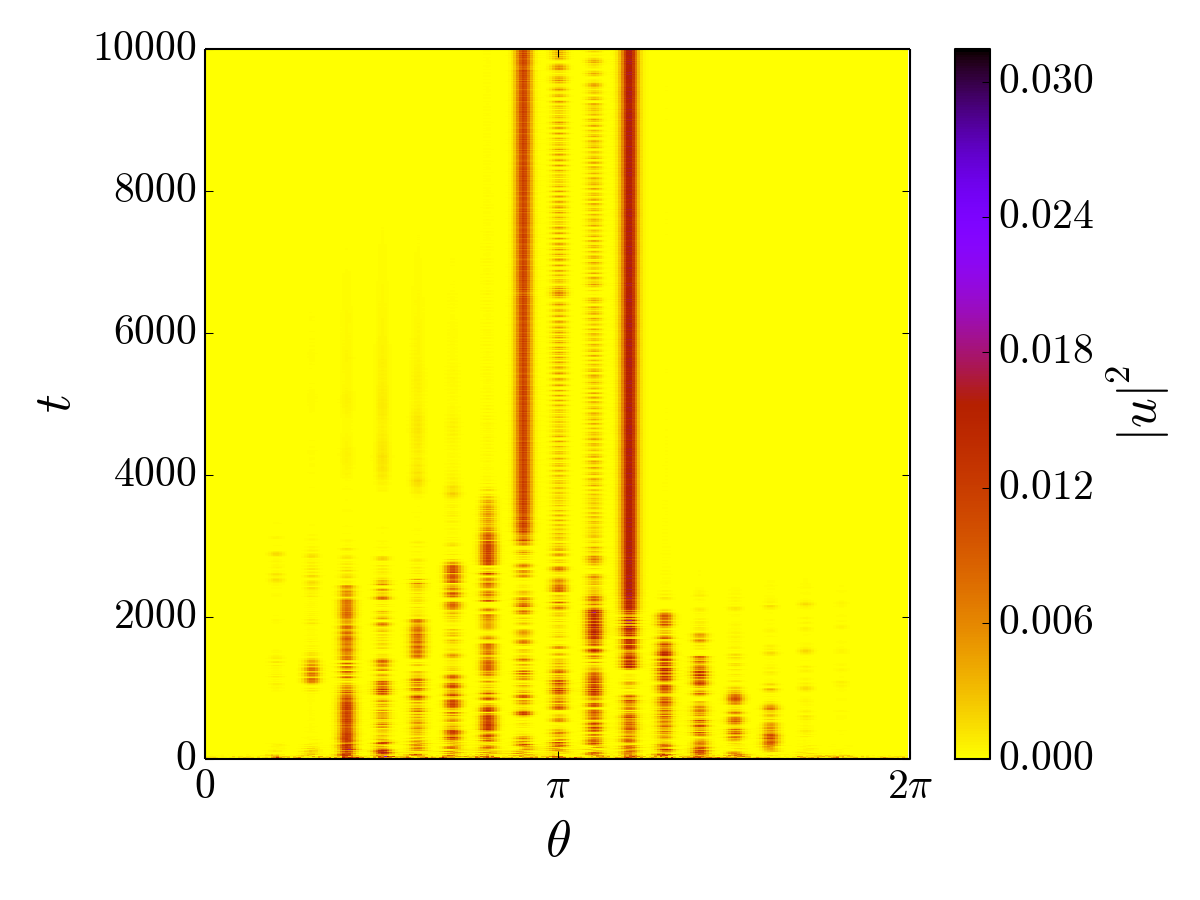}
\caption{\color{black}(Color online) Space-time evolution of atomic density $u(x,t)$ with $\beta=50$ in the presence of localized dissipations.  The initial condition is that of a "flat" equal amplitude wavefunction 
with random phases.
  \label{densflat}}
\end{figure}

\begin{figure}[htb!]
\includegraphics[angle=0.0,clip,width=\columnwidth]{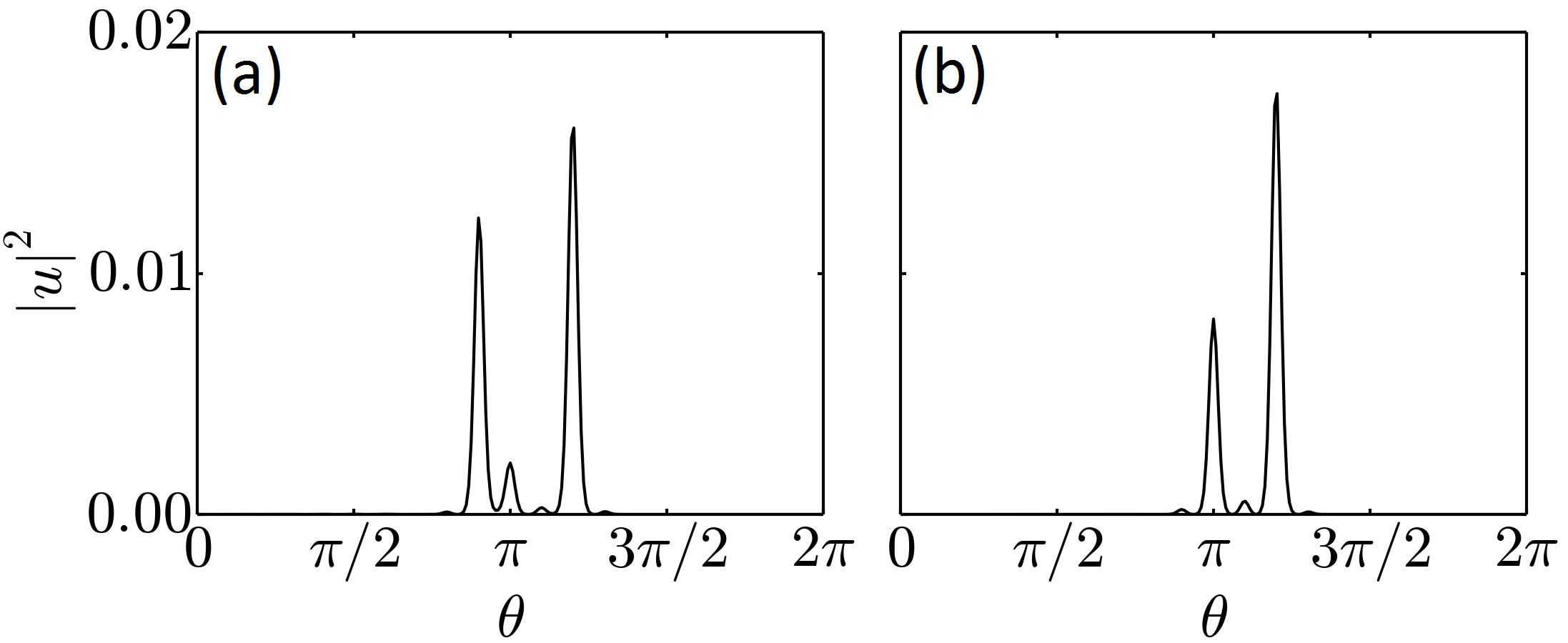}
\caption{\color{black}(Color online) Intensity distribution of localized solution obtained from applying 
dissipation to an initially flat wavefunction at $t=10000$ (a) and $t=100000$ (b). 
  \label{flat}}
\end{figure}

For completeness we show that localized structures can also be obtained via localized dissipations by starting from a homogeneous distribution of atoms across the optical lattice with random phases in analogy with what has been done in the DNLSE \cite{livi06,franzosi11}. In the example here, we first run a transient without dissipations for 1000 time-steps. After this, dissipation is turned on as shown in Fig.~\ref{densflat}. There is a first localization to two peaks, (see Fig. ~\ref{flat}(a)). The amplitudes of the peaks fluctuate and eventually, at long time scales (around $t=35000$), the peaks move closer to each other so that only one potential well separates them (see Fig.~\ref{flat}(b)). To observe this behavior the nonlinearity has been increased to $\beta=50$.

\section{Traveling Localized Dissipations}
\label{travelling}

\begin{figure}[htb!]
\includegraphics[angle=0.0,clip,width=\columnwidth]{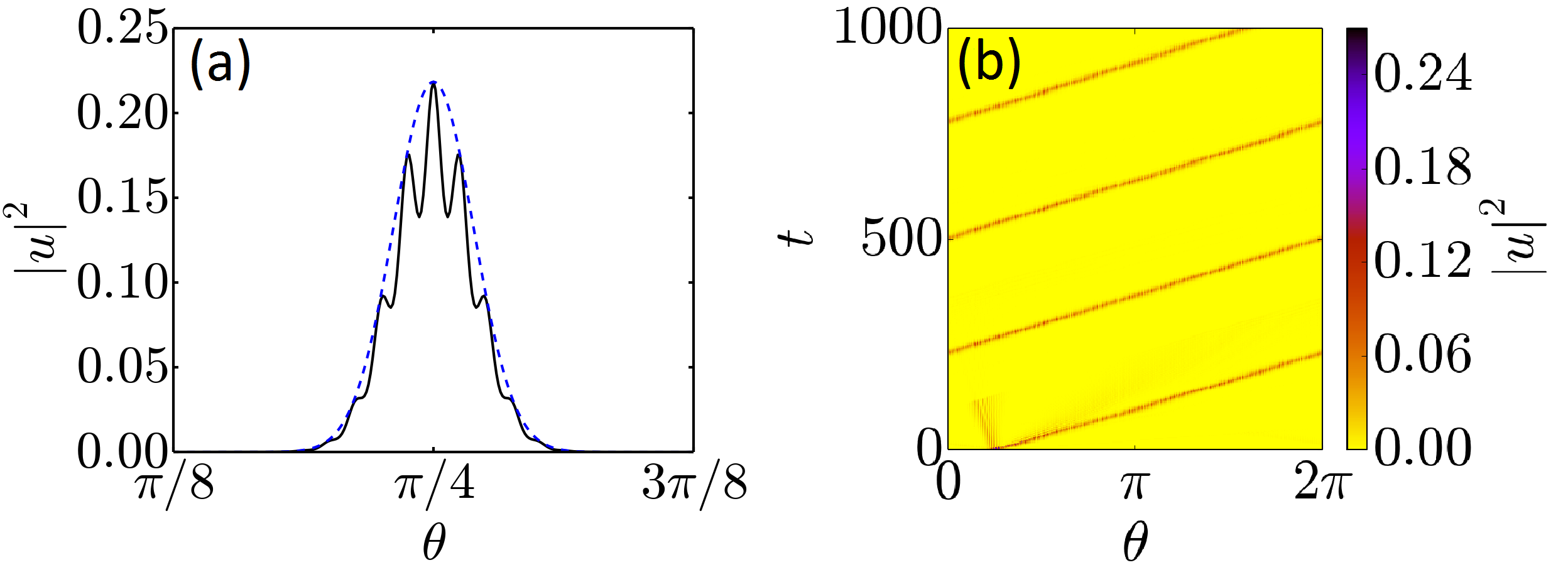}
\caption{\color{black}(Color online)
(a) Initial condition for the formation of a TLS. Note that this distribution would be an ordinary Gaussian shape in the DNLSE (blue dashed line) (b) Space-time evolution of atomic density $u(x,t)$ of the TLS with $\Lambda=1$ and dissipations $\rho = 0.5$. 
  \label{travelnonlin1}}
\end{figure}

By using an initial Gaussian wavepacket with an additional momentum, traveling breathers can be formed in the DNLSE \cite{hennig16,trombettoni01,franzosi11}. In order to simulate this procedure in the continuous case and stabilise a traveling lattice soliton (TLS), we have used an initial distribution made of a "Gaussian of Gaussians" (see Fig. ~\ref{travelnonlin1} (a)). In the DNLSE where each potential well corresponds to a single lattice point, our distribution reduces to a normal Gaussian shape (see dashed line in Fig. ~\ref{travelnonlin1} (a)). With the addition of an initial momentum $p$ (here set to $cos(p)=-0.95$), a traveling peak is formed in the continuous model. We then apply dissipation in the angular position opposite to this peak in a way similar to what is described in \cite{franzosi11} for the DNLSE. Since the atomic density peak is traveling, the point at which dissipation is applied also moves.

\begin{figure}[htb!]
\includegraphics[angle=0.0,clip,width=\columnwidth]{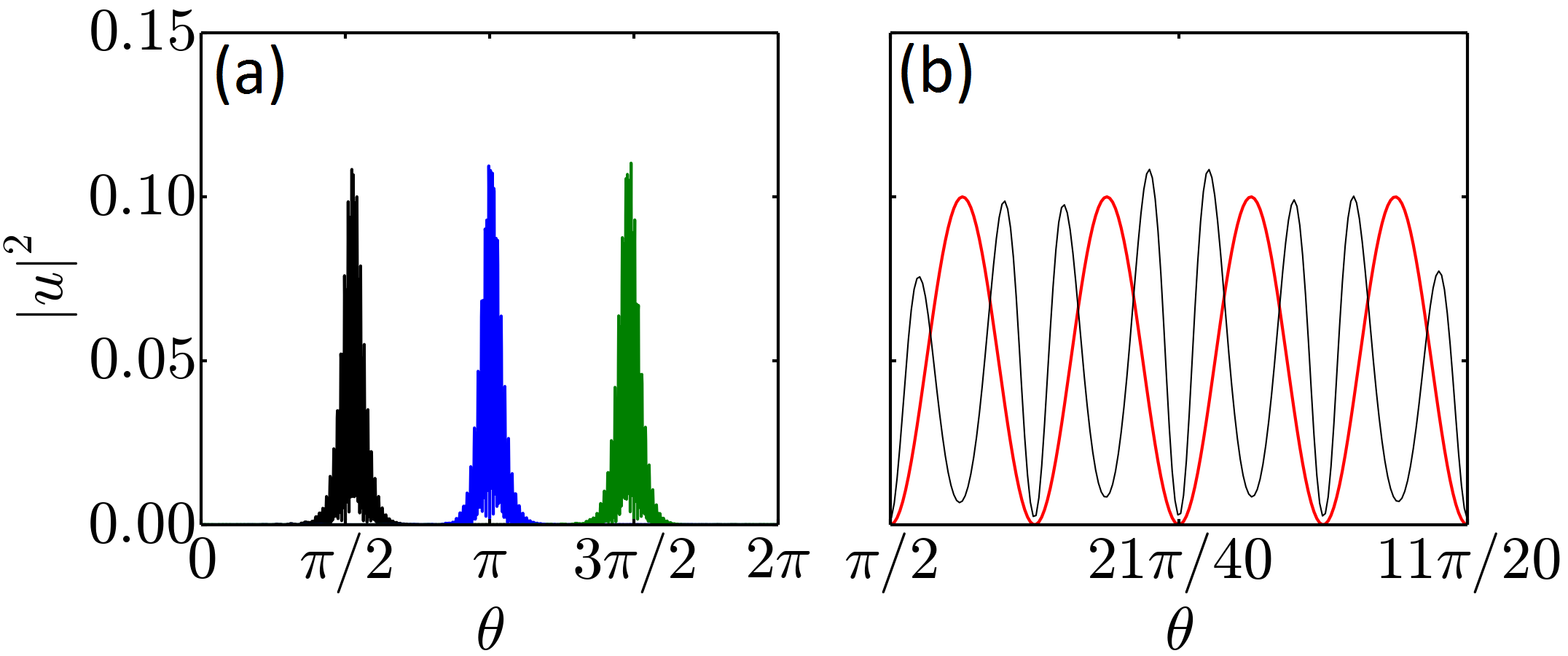}
\caption{\color{black}(Color online)
(a) Intensity distribution of a TLS at $t=10150$ (black), $t=10220$ (blue) and $t=10290$ 
(b) Close-up of intensity distribution at $t=10150$ (black, thin) with the periodic potential (red, thick), 
showing the two peaks-per-potential well.
  \label{travelnonlin1intens}}
\end{figure}

In the example shown in Figures~\ref{travelnonlin1} and \ref{travelnonlin1intens}, we consider $\beta=1.0$, $V_0=10$, and $cos(p)=-0.95$ and dissipations given by $\rho=0.5$ over 4 lattice wells. At the begining of the simulation, a certain amount of atoms remains stationary after the traveling peak is formed. This can be seen in Fig.~\ref{travelnonlin1} (b), with the high amplitude stationary part of the wavefunction visible until $t\approx 100$, when these atoms are removed from the lattice by the moving dissipation beam. At long time scales, the peak shapes into a TLS that travels at a constant speed (shown in Fig.~\ref{travelnonlin1intens}(a)). It is important to note that without dissipations, the atoms that do not travel with the moving peak eventually spread across the lattice, giving rise to a large background noise. As the moving peak travels and interacts with the background, its amplitude reduces since it loses atoms to the background. By $t\approx 1600$, the height has decreased by half and by $t\approx 3000$ the conservative traveling peak has disappeared. In contrast in the presence of the moving dissipation, the TLS survives on much longer time scales, maintaining the same height after $t\approx 40000$. The fact that dissipation helps instead of hinder the formation of a TLS is even more surprising since, at difference with the stationary lattice solitons, TLS require the presence of a background in order to overcome the unavoidable Peierls–-Nabarro barriers \cite{kivshar93,franzosi11}. The presence of the localized dissipation is then twofold: on one side it removes enough stationary background noise to help with the localization of the TLS and on the other it moves with the traveling background thus maintaining it to the level necessary for the motion and stability of the TLS.

\begin{figure}[htb!]
\includegraphics[angle=0.0,clip,width=\columnwidth]{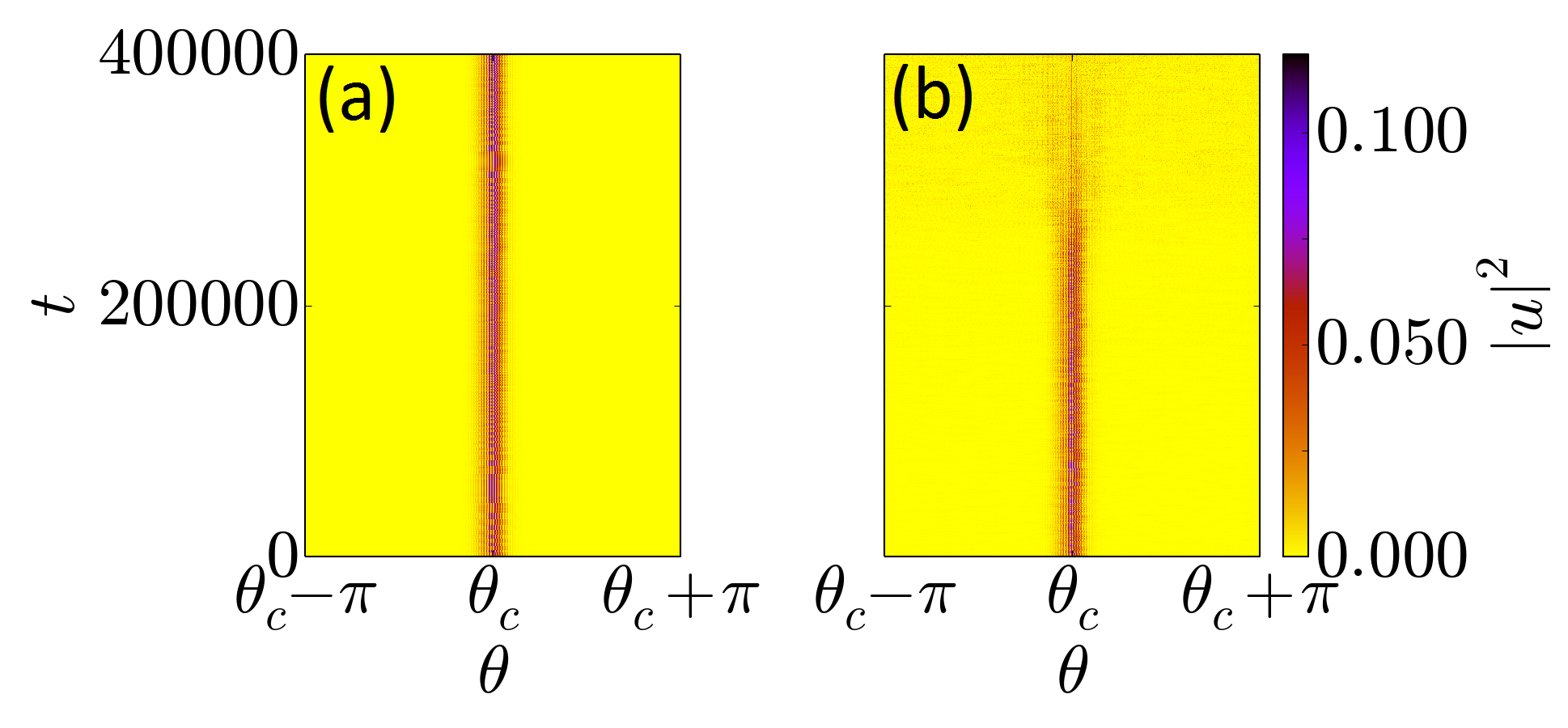}
\caption{\color{black}(Color online)
Temporal evolution of the intensity distribution of the TLS initiated via (\ref{twoptls}) for the case with localized dissipations ($\rho=-0.4$) (a) and without localized dissipations ($\rho=0$) (b). Note that TLS is traveling along the ring but each distribution has been shifted so to have the TLS maximum at the same angular location).
  \label{dissvscons}}
\end{figure}

It is important to note that the TLS of Fig.~\ref{travelnonlin1intens}(a) formed via the localized dissipation is a 'higher-order' TLS with two atomic density peaks per potential well (see Fig.~\ref{travelnonlin1intens}(b)). Due to its shape, this TLS has no counterpart in the DNLSE. We have determined an approximate form of the amplitude of the TLS displayed in Fig.~\ref{travelnonlin1intens} that can be used as initial conition at time $t=0$ and given by
\begin{eqnarray}
u ( \theta) &=& - 7.66 \;\; A  \exp \left[ i \frac{p M (\theta-\pi)}{2\pi} \right] \nonumber \\
& &\sin \left[ M (\theta-\pi) \right ] \mathrm{sech} \left[ AM(\theta-\pi) \right]
\label{twoptls}   
\end{eqnarray}
where $A$  is a parameter that depends on the width of the TLS. For $A=1/(7.5 \pi)$ and $p=-0.4$ we obtain a fit of the TLS in Fig.~\ref{travelnonlin1intens} as accurate as few percents. Having determined the approximate TLS shape in Eq. (\ref{twoptls}), one can use it as an initial condition for the formation of the double peak TLS in the presence or absence of dissipations. With dissipations $\rho=0.5$, we have verified that the TLS of Fig.~\ref{travelnonlin1intens} forms much faster when using the wavepacket (\ref{twoptls}) as intial condition instead of the Gaussian wavepacket. Figure \ref{dissvscons} (a) shows that this TLS survives for extremely long time scales with an extremely small loss of atomic density or energy. The steady loss due to dissipation is so small that after one million time units, the atomic density only decreases by 0.21\%. This is similar to what happens to the stationary lattice solitons in Section \ref{stationary} when boundary losses approached irrelevance at the tails of the lattice soliton.

Without localized dissipations, a traveling peak starting from (\ref{twoptls}) survives for a long time (see Fig. \ref{dissvscons} (b)). However, in the absence of dissipations, the background noise eventually grows and absorbs the peak as shown in the last stages of Fig. \ref{dissvscons}(b). These features demonstrate that localized dissipations are necessary for both the formation and the stability of the double peak TLS when starting from wavepacket distribtuions of atoms in the lattice with a given momentum. 

\begin{figure}[htb!]
\includegraphics[angle=0.0,clip,width=\linewidth]{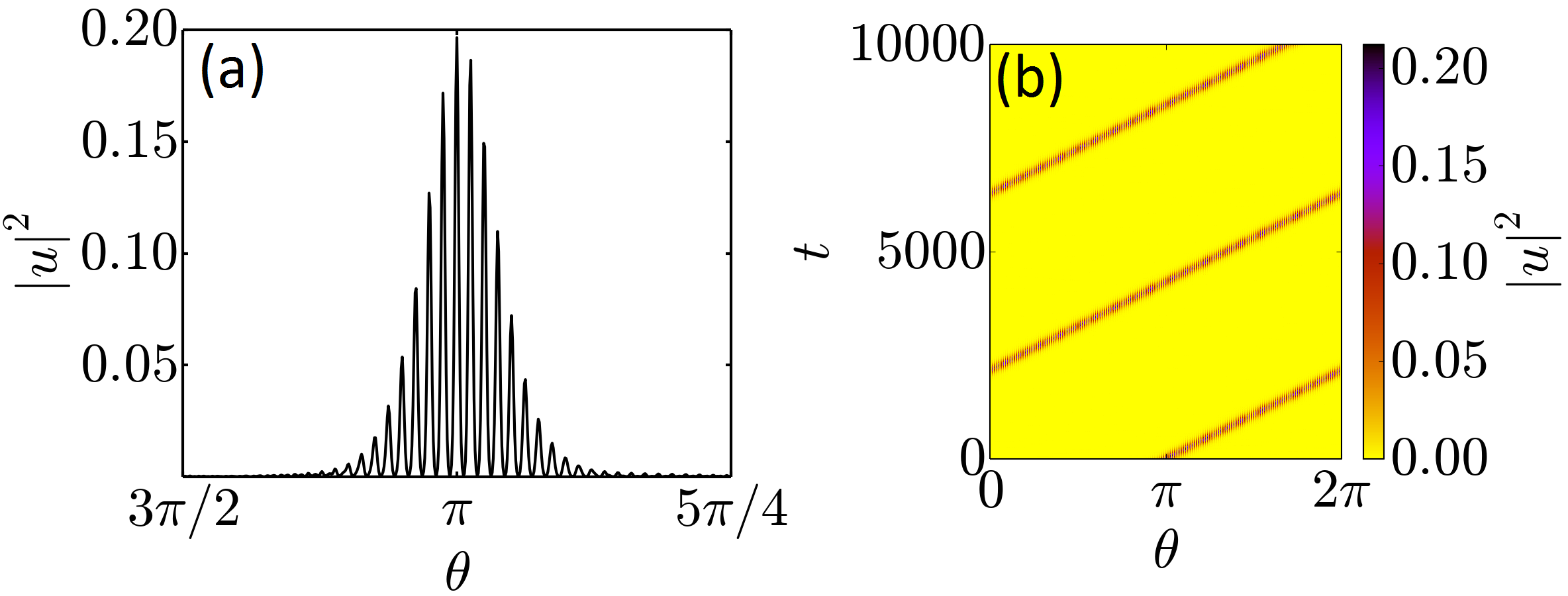}
\caption{\color{black}(Color online) TLS with one-peak stabilized by loclaized dissipations.
  \label{traveldens1p}}
\end{figure}

We have also applied localized traveling dissipation to TLS with one peak per potential well by using the analytical approximation of \cite{sakaguchi04} 

\begin{eqnarray}
u ( \theta) &=& 8.11 \;\; A  \exp \left[ i \frac{p M (\theta-\pi)}{2\pi} \right] \nonumber \\
& &\cos \left[ \frac{M (\theta-\pi)}{2} \right ] \mathrm{sech} \left[ AM(\theta-\pi) \right]
  \label{malomedtravel}
\end{eqnarray}
with $A$ and $p$ being the amplitude and the momentum of the TLS respectively.  In Fig.~\ref{traveldens1p} (a) and (b) we set to $A=0.3/(2\pi)$ and $p=-0.5$ and show the amplitude of the initial condition (\ref{malomedtravel}) and its temporal evolution in the ring, respectively. It is important to note that with or without dissipation, the inital condition (\ref{malomedtravel}) quickly develops a noisy background on which the TLS travels while remaining well approximated by (\ref{malomedtravel}) in the potential wells where atomic localization takes place. The dissipation clears up stationary noise, but does not destroy the TLS with one peak per potential well.  The atomic density is only slightly affected by the dissipation, which decreases by $\sim0.12 \%$ after one million time units, even slower than the higher-order TLS.

\section{Collision of a Traveling and a Stationary Lattice Soliton}
\label{collisionTSb}

In this section we investigate the collision of the TLS with two peaks per potential well  (previously stabilized by the localized dissipations) and a stationary lattice soliton generated with the same method discussed in Section \ref{stationary}. The height of the stationary soliton is varied by changing the width of the initial Gaussian wavepacket via a modification of the $\gamma$ parameter. 

\begin{figure}[htb!]
\centering
\includegraphics[angle=0.0,clip,width=\columnwidth]{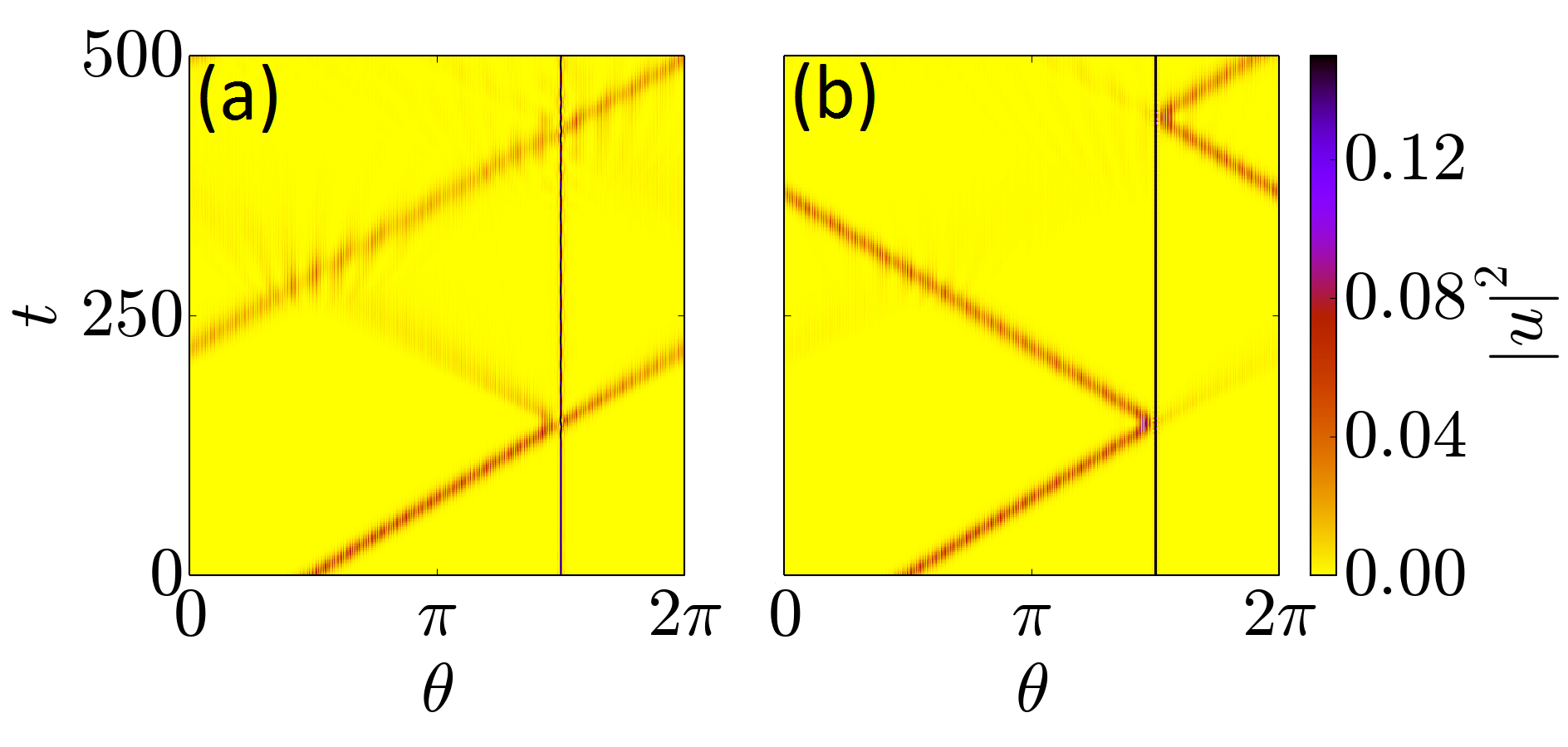}
\caption{\color{black}(Color online) 
A TLS colliding with a SLS of amplitude $\approx 0.45$ (a) and $\approx 0.95$ (b).
  \label{collision}}
\end{figure}

In Fig.~\ref{collision} the temporal evolution of the atomic density of both lattice solitons at successive collisions in the ring is displayed for zero dissipations. The TLS and the SLS are initially as far apart in the ring from each other as possible. The amount of atomic density that passes through the stationary lattice soliton at each collision is determined by its height. The higher the stationary lattice soliton, the less atomic density passes through, as shown in the examples of Fig.~\ref{collision}. For example, when the amplitude of the SLS is low (see Fig.~\ref{collision} (a)), the majority of the atomic density in the TLS passes through the stationary one at the point of collision with only a small amount being reflected. After each collision, the atomic density that has been reflected interferes with and scatters the atomic density of the TLS that has been transmitted by the SLS. This makes the TLS weaker and weaker as time goes on.  

When the amplitude of the SLS is high ($\approx 0.95$ in Fig.~\ref{collision} (b)), the majority of the atoms in the TLS reflects off of the stationary one while only a small amount manages to tunnel through. The small amount of atomic density that tunnels through appears to have no major effect on the reflected TLS, which manages to survive longer than in the previous example. 

\section{Collisions of Two Traveling Lattice Solitons}
\label{collisionTT}

\begin{figure*}[t]
\centering
\includegraphics[angle=0.0,clip,width=2\columnwidth]{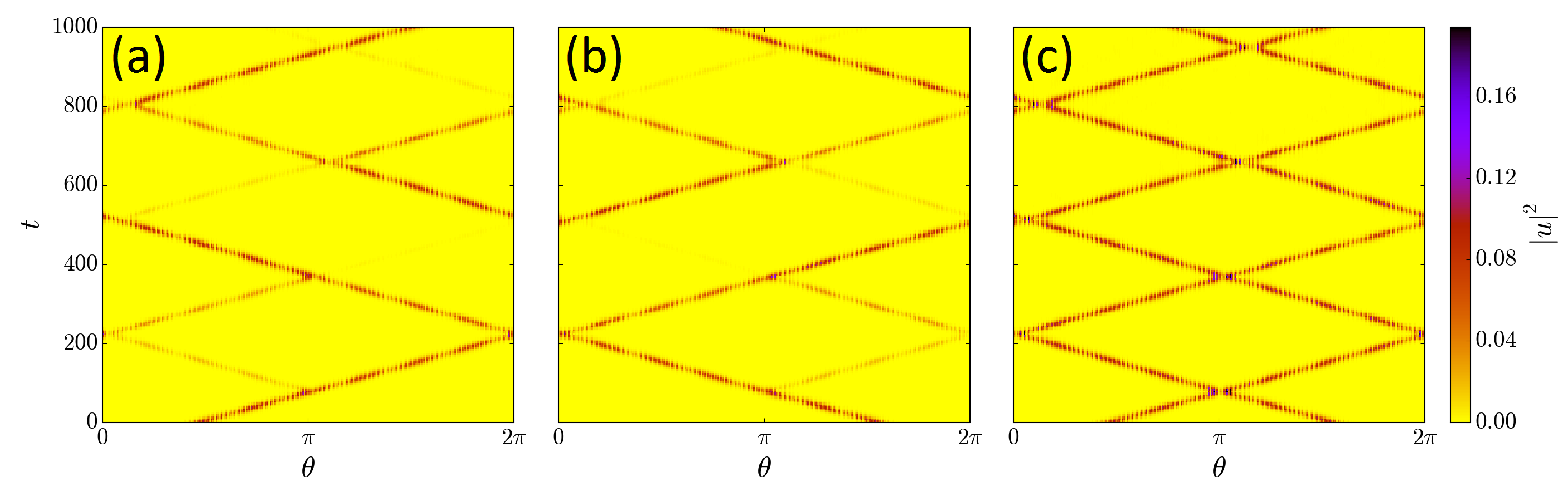}
\caption{\color{black}(Color online) Collision of two TLS with two peaks per potential well. The total atomic density profile of the collisions is shown in (a) while the atomic density profile from each initial TLS is plotted in (b) and (c).
  \label{collision2peak}}
\end{figure*}

For completeness, we examine the collisions of two TBS circling in the ring. In the first example, in  Fig.~\ref{collision2peak}, we use the TBS with two peaks per potential well as described in Section  \ref{travelling} and with $\beta=1$ and $V_0=10$. We first position two identical TSB at opposite sides of the ring ($\approx \pi$ radians apart), make them traveling in the opposite directions ($p=0.5$ and $p=-0.5$, respectively) and then make them collide. Since dissipations would interfere with the process of collisons, we set $\rho=0$ for both TLS. As demonstrated in Fig. \ref{dissvscons} (b), the TLS with no dissipations survives for a long time during which more than a hundred collisions can take place. We focus here on the first couple of collisions to establish the nature of the interaction of the TLS at short distances and for interferometric properties. The collision from the two TLS results in two seemingly identical TLS at the output (see Fig.~\ref{collision2peak} (a)). We have verified that both atomic density and energy have not changed in each of the output TLS with respect to the input. 

In order to find out if the TLS have gone through one another or have reflected each other, we have split the wavefunction in two by substituting $u=u_1+u_2$, where $u_1$ represents the atoms of one TLS and $u_2$ in the other, into Eq.~(\ref{GPEnorm}) to get:

\begin{eqnarray}
  \label{split1}
  i \frac{\partial u_1(\theta,t )}{\partial t} = \left(-\frac{\pi^2}{2M^2}\frac{\partial^2}{\partial \theta^2} + V_0 sin^2 \left(\frac{M \theta}{2}\right) 
  + \beta |u_1+u_2|^2\right) u_1 \nonumber \\
  \label{split2}
i \frac{\partial u_2(\theta,t )}{\partial t} = \left(-\frac{\pi^2}{2M^2}\frac{\partial^2}{\partial \theta^2} + V_0 sin^2 \left(\frac{M \theta}{2}\right) 
  + \beta |u_1+u_2|^2\right) u_2 \,
  \label{u1plusu2}
\end{eqnarray}
We find that, when the TLS collide, some of the atomic density from each TLS pass through the other while the remaining part is reflected.  When this happens, the reflected atomic density of each TLS merges with the transmitted part of the other one. This happens in such a way that the two TLS that result from the collision have approximately the same shape as the original ones, despite containing a mixture of the atomic density from each of them. We have verified that the results of the numerical simulations of Eqs. (\ref{u1plusu2}) reproduce exaclty those of the simulations of Eq. (\ref{GPEnorm}) when cosidering $u=u_1+u_2$. In this particular example, $\sim79.5\%$ of the atomic density of each TLS passes through the other one at each collision. The evolution of the atomic density distributions of each initial TLS are plotted in Fig.~\ref{collision2peak} (b) and (c) respectively, showing how each TLS splits at each collision. The transmitted/reflected fractions of atomic density of the two TLS in the collisions does not change when starting the collision process from a different intial lcoation of the TLS. However, we have measured that these fractions change with the depth of the lattice potential as reported in Table ~\ref{collisiontable2peak}. 

  \begin{table}
    {\color{black}
    \caption{Percentage of atomic density reflected and transmitted in collisions between 2 higher-order TLS \label{2peaktable}}
    \label{collisiontable2peak}
    \begin{tabular}{c c c c c c c c}
      \hline
      \hline
      $V_0$ & reflection & transmission\\
      \hline
      7.0 & 11.8 & 88.4\\
      7.5 & 12.6 & 87.4\\
      8.0 & 13.7 & 86.3\\
      8.5& 15.1 & 85.0\\
      9.0& 16.8 & 83.2 \\
      9.5 & 18.6 & 81.5\\
      10.0 & 20.4 & 79.6\\
      \hline
    \end{tabular}}
  \end{table}

\begin{figure*}[t]
\centering
\includegraphics[angle=0.0,clip,width=2\columnwidth]{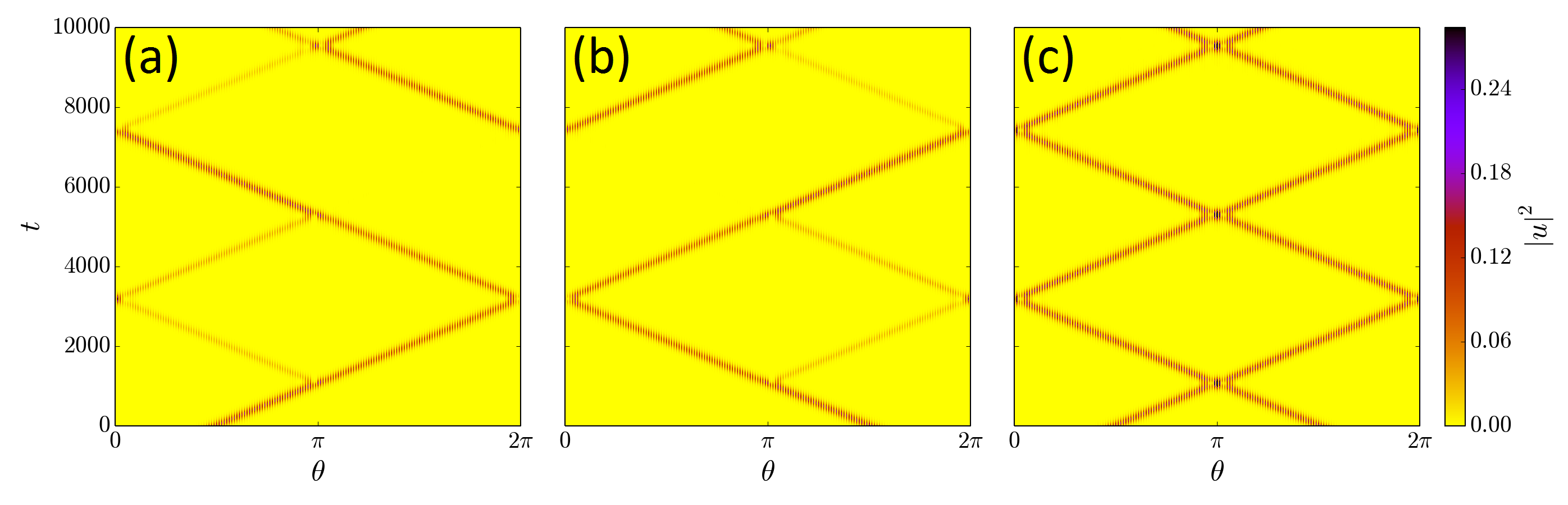}
\caption{\color{black}(Color online) Collision of two TLS with a single peak per potential well. The total atomic density profile of the collisions is shown in (a) while the atomic density profile from each initial TLS is plotted in (b) and (c).
  \label{collision1peak}}
\end{figure*}

Similar results of collisons occurs with the TLS with just one peak per potential well. In Fig.~\ref{collision1peak},  we show collisions of these TLS for $V_0=10$ and $\beta=0.041$.  Again, the TLS ``swap'' atomic density at each collision, with the shape of the resulting TLS largely unchanged. Here, $\sim 77.0\%$ of the atomic density in each TLS stay with the ``original'' one at each collision, while the rest join the other ones. In Table ~\ref{collisiontable1peak} we show the dependence of the transmitted/reflected fractions of atomic density in the collisions of TLS with a single peak per potential well when changing the depth of the optical lattice.

  \begin{table}
    {\color{black}
    \caption{Percentage of atomic density reflected and transmitted in collisions between 2 TLS \label{1peaktable}}
    \label{collisiontable1peak}
    \begin{tabular}{c c c c c c c c}
      \hline
      \hline
      $V_0$ & reflection & transmission\\
      \hline
      9.0 & 14.7 & 85.3\\
      9.5 & 18.3 & 81.7\\
      10.0 & 23.0 & 77.0\\
      10.5& 29.5 & 70.5\\
      11.0& 38.2 & 61.8 \\
      \hline
    \end{tabular}}
  \end{table}

\section{Conclusions}
We have analyzed the effect of local dissipation on BEC in a ring lattice.  We found that the dissipation can both generate and stabilize stationary and traveling lattice sollitons (SLS and TLS, respectively).  A TLS with two intensity peaks per potential well was introduced that does not have a counterpart in the discrete NLS.  This can be generated, via an initial Gaussian wavepacket (as in the discrete model) with dissipation.  This does not survive without losses in the long term. We then investigated the collisions of this TLS with different SLSs and found that the interaction and survival of the TLS depends on the amplitude of the SLS.  We also analyzed the collisions of two TLS in the ring.  We found that some of the atoms in each TLS merge with the colliding one while some are reflected in such a way that the shape of the resulting TLS' intensities stays the same.  This collisional property depends on the potential depth of the lattice.  The amount of atoms that are transmitted (reflected) during the collision is smaller (larger) in deeper lattices and larger (smaller) in shallower lattices.

A possible application of the TLS in a ring lattice is interferometry. The TLS can collide with extra potential barriers added to the lattice.  This has been proposed for attractive BEC without a lattice in \cite{helm12,helm14,helm15}.  With an optical lattice, there is the possiblilty of the interferometric features, such as Sagnac effects, to work with a repulsive BEC and with higher order TLS.  

The SLS and TLS solutions obtained via localized dissipations are robust determinisitc features to small fluctuations. It should also be noted that although the model and equations of this paper have been used to describe the situation of BEC in a ring lattice, they can also be generalized to light propagating in cylindrical arrays of waveguides.

\section{Acknowledgements}
R. C. acknowledges support from an EPSRC DTA Grant No. EP/M506643/1.

\end{document}